\documentstyle[epsf,aps,prl]{revtex}

\epsfverbosetrue
\newcommand{\figwidth}{\columnwidth}
\renewcommand{\epsfsize}[2]{\figwidth}

\newlength \figwold
\newcommand{\epscenter}[2]
   {\figwold=\figwidth \figwidth=#1\hfill\epsfbox{#2}\hfill\figwidth=\figwold}

\reversemarginpar

\begin{document}

\title{Molecular Dissociation in Hot, Dense Hydrogen}
\author{W. R. Magro$^1$, D. M. Ceperley$^2$, C. Pierleoni$^3$, and B. Bernu$^4$}
\address{
$^1$Theory Center and Laboratory of Atomic and Solid State Physics,
Cornell University, Ithaca, NY 14853
\\
$^2$National Center for Supercomputing Applications and
Department of Physics, University of Illinois at Urbana-Champaign,
Urbana, Illinois 61801
\\
$^3$Dipartimento di Fisica, Universita de L'Aquila I-67100 L'Aquila (Italy)
and INFM, sezione di Roma I, 00185 Roma
\\
$^4$Laboratoire de Physique Theorique des Liquides,
Universit\'e P. et M. Curie, 75252 Paris Cedex 05
}
\date{Received \today}
\maketitle
\begin{abstract}
We present a path-integral Monte Carlo study of
dissociation in dense hydrogen ($1.75 \leq r_s \leq 2.2$, with $r_s$
the Wigner sphere radius).
As temperature is lowered from $10^5$ to 5000 K, a molecular hydrogen
gas forms spontaneously
from a neutral system of protons and electrons.
At high density, $r_s < 2.0$, thermally activated dissociation is
accompanied by decreasing pressure, signaling the presence
of a first order transition and critical point.
The decrease in electron kinetic energy during dissociation
is responsible for the pressure decrease and transition.
At lower density the phase transition disappears.
\end{abstract}
\pacs{PACS numbers: 05.30.-d,02.70.Lq, 64.30.+t, 71.10.-w}
\narrowtext
\twocolumn

Despite its simple composition, hydrogen is a complex substance with a rich phase diagram.
While recent attention has been directed primarily at its ground state
properties, much of its behavior at high temperature and density,
particularly important in astrophysics,
remains uncertain.
Molecular dissociation can occur thermally or through compression, as
the electrons are forced into high energy states by the increasing
chemical potential.
A fundamental question is whether hydrogen crosses a phase boundary 
as the dense molecular fluid transforms into
a fully ionized plasma.
The answer is not simple to deduce, since dissociation
occurs in a region where both thermal and degeneracy effects, as well
as many-body quantum effects, are important.
Further, as in a liquid-gas transition, there is no change in symmetry.
If such a ``plasma phase transition'', as it has come to be known,
indeed exists, it remains unclear whether
the dissociation and ionization processes occur simultaneously
or if instead there exists an intermediate atomic-like state.

Direct observation of the dissociation process is hindered by the high
pressures and temperatures required.
Recent shock-wave compression measurements, the most promising
experimental approach, see no evidence for a first order
phase transition~\cite{nellis95}.

One theoretical approach to these questions has been to separately
model the molecular and metallic phases, then equate the Gibbs
free energies to locate a transition.
More advanced models generally include more chemical species, such
as ${\rm H_2}$, H, protons, and electrons.
A particularly sophisticated chemical treatment, the free energy
model of Saumon and Chabrier~\cite{saumon92}, predicts a first order
phase transition from the molecular fluid to a partially
ionized atomic gas.
The density discontinuity associated with this phase transition would 
sharply alter current estimates of the interior mass distribution
of the giant planets.
Chabrier, {\em et al.\/}\ have shown that a plasma phase transition is in fact 
{\em required\/}
to obtain agreement between the most sophisticated structural models of
Saturn and the measured gravitational moments~\cite{chabrier92}.

At such elevated temperatures and densities, it is unclear whether
a chemical picture is adequate, since the very chemical species
in the system are evolving and the relevance of the terms `atom'
and `molecule' is uncertain.
The restricted path-integral Monte Carlo (RPIMC) method is unique
in its ability to simulate fully interacting many-fermion
quantum systems in thermodynamic equilibrium with a
minimum of approximations.
In particular, hydrogen can be modeled within RPIMC as
a collection of fully interacting electrons and protons.
This relatively new method has been previously applied only
to $^3$He~\cite{ceperley92},
isotopic helium mixtures~\cite{boninsegni95},
and the dense hydrogen plasma~\cite{pierleoni94}.
We apply it here to investigate the nature of
molecular dissociation in dense hydrogen, thus circumventing the
problems associated with chemical models.

We model hydrogen as a neutral mixture of 32 protons
and 32 unpolarized electrons in a periodically repeated cubic cell
and in equilibrium at a temperature, $T=1/k_B\beta$.
Density is specified in terms of the Wigner sphere radius, $r_s$, defined by
${4\pi/3}(r_s a_0)^3 \equiv n^{-1}$,
where $a_0$ is the Bohr radius and $n$ is the average electron density.
We use the fully interacting, non-relativistic Hamiltonian for this system.
The density matrix, $\rho(\beta)\equiv e^{-\beta {\cal{H}}}$,
contains complete thermodynamic information about the system with
observables given as:
\begin{equation}
\label{eq:observable}
\left\langle{\cal{O}}\right\rangle = {{\rm Tr}\left[ \cal{O}\rho(\beta) \right]
\over {{\rm Tr}\left[ \rho(\beta) \right]}}
= {\int dR \, \left\langle{R|\cal{O} \rho(\beta)|R}\right\rangle \over
\int dR \, \left\langle{R|\rho(\beta)|R}\right\rangle},
\end{equation}
where $R\equiv\{{\bf r}_1,\ldots,{\bf r}_N\}$ specifies a configuration of the $N$ particles.
Due to its exponential form, the density matrix can be factored as
$\rho(\beta) = {\left [{\rho(\tau)}\right ]}^M$ if $M\equiv\beta/\tau$.
Eq.~(\ref{eq:observable}) then becomes
a path-integral which is well-suited for Monte Carlo
evaluation using a multi-stage Metropolis algorithm~\cite{ceperley95}.
The problem is thus reduced to one of evaluating off-diagonal elements
of the high-temperature density matrix, $\rho(\tau)$.

To develop an expression for $\rho(\tau)$, we first split
the Coulomb potential into one short- and one long-ranged term.
The high-temperature density matrix, $\rho_s(R,R';\tau)$, in the
absence of the long-range potential is written as a product of
pairwise terms, which are computed with a 
matrix-squaring technique~\cite{storer70,ceperley95}.
Next, we introduce the long-range interaction perturbatively and write
\begin{equation}
\label{eq:sep-rho}
\rho(R,R';\tau) = \rho_s(R,R';\tau) e^{-{1\over2}\left [{U(R;\tau)+U(R';\tau)}\right ]}
\end{equation}
with the long-range action, $U(R;\tau)$, determined using the
random phase approximation~\cite{pines89}
(details are given in Ref.~\cite{magro:thesis}).
Evaluation of the path-integral requires a non-zero
$\tau$, which introduces a systematic `time-step error,' which
decreases with decreasing $\tau$.
In the present work $\tau^{-1} = 10^6\,k_B{\rm K}$,
which gives a reasonable tradeoff between computational effort and accuracy.
At this value, the total energy is accurate to about $0.06\,{\rm Ry}$ or $5\%$ per atom.

In the path-integral formulation, electrons and protons
are put on equal footing from a quantum-mechanical point of view,
since each particle is fully represented by a Feynman path.
This causes no particular difficulty, in contrast to other methods,
such as local density functional calculations.
The Fermi temperatures of the protons and electrons at the highest density
considered, $r_s=1.75$, are $103\,{\rm K}$ and $190000\,{\rm K}$, respectively.
While the Fermi character of the protons is negligible under the conditions
of this study, the effects of electron exchange are substantial.
Without Pauli repulsion, a proton-electron mixture is thermodynamically
unstable\cite{theilhaber91}.
The density matrix must be
anti-symmetric under exchange of spin-like electrons,
but a direct anti-symmetrization procedure will be statistically
very inefficient
due to cancelation of negative and positive terms.
This is the well-known {\em sign problem\/} of fermion Monte Carlo techniques.
We circumvent this problem by using the fixed-node approximation, in which
the paths are restricted to lie within
a set of physically motivated trial nodes~\cite{ceperley91,ceperley92}.
The procedure becomes exact when the trial nodes coincide with the exact
fermion nodes.
We use the nodes of the free-particle density matrix.
While these nodes become exact only at high temperature,
they capture more of the physics than one might expect.
Their merits are discussed in more detail in Ref.~\cite{ceperley92}.
In the absence of exact results,
the errors due to incorrect trial nodes can only be estimated from
analogous ground state calculations~\cite{natoli93}.
These show that for reasonable nodes,
the energy is largely insensitive to exact nodal positions.
Typical fixed-node errors at $T=0$ are about $0.004\,{\rm Ry}$ per
atom, and the error at finite temperature should be smaller.

Dense hydrogen is therefore not only an interesting problem, but also an
important test case for the method.
Previously, we successfully applied the same method to a dense hydrogen
plasma, obtaining good agreement with theoretical
predictions~\cite{pierleoni94}.
We have also tested our program on
the isolated hydrogen atom, hydrogen
molecule, and helium atom.

In brief, our results for dense hydrogen near dissociation generally
support the findings of Saumon and Chabrier from their
chemical model~\cite{saumon92}, though the quantitative details differ.
Most significantly, both approaches find behaviors consistent
with and suggestive of a first order plasma phase transition.
As in the chemical model, our molecular gas dissociates first into a partially
ionized atomic-like fluid, then gradually transforms into an ionized plasma.

At the lowest temperatures and densities considered
$\left(r_s=\left\{1.75, 1.86, 2.0, 2.2\right\}, T=5000\,{\rm K}\right)$,
a molecular hydrogen gas forms, with
the bond somewhat contracted from its free space length,
0.742 \AA.
At $r_s=2.2$ the bond length is 0.67 \AA\ and further decreases with increasing
density, reaching 0.65 \AA\ at $r_s=1.75$.
The bond contraction is nearly temperature-independent and apparently results
from a stiff effective intermolecular repulsion.
This repulsion also
leads to an excluded region surrounding each molecule.
From $g_{pp}(r)$, the proton-proton pair distribution,
at ($r_s=2.2$, $T=5000$ K) we estimate the radius of the
repulsive core to be $0.6$ \AA, somewhat smaller than previous
estimates~\cite{saumon92}.

\begin{figure}
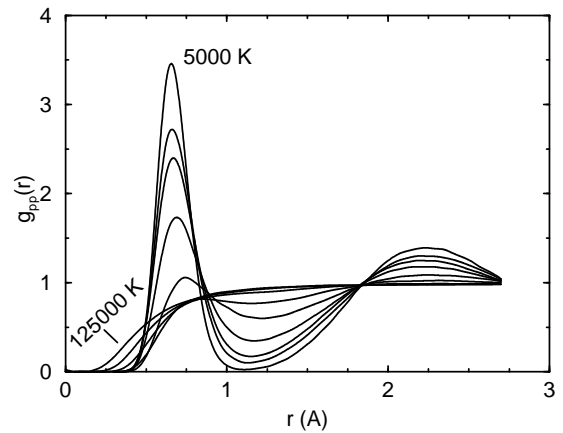

\epscenter{\figwidth}{fig1}
\caption
{\label{fig:grpp}
Proton-proton pair correlation functions of hydrogen at $r_s=2.0$.
Temperatures shown are 5000, 6944, 7813, 8927, 10000, 12500, 15625,
21250, and 125000 K.
As temperature is lowered, a molecular hydrogen gas forms with
bond length slightly contracted from the free space value.
}
\end{figure}

\begin{figure}
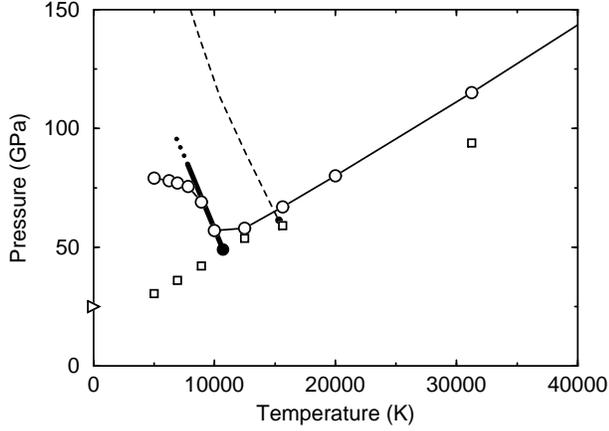

\epscenter{\figwidth}{fig2}
\caption
{
\label{fig:pressure}
Computed pressures (open circles) in the present work and
from the chemical picture~\protect\cite{chabrier94:private}
(squares) for $r_s=2.0$.
Statistical errors are smaller than the symbol size.
The dashed line is the phase coexistence line transition proposed by Saumon
and Chabrier~\protect\cite{saumon92}.
Our estimate of the same line is given by the thick solid line.
The ellipses ($\cdots$) indicate the line continues to higher pressure.
The triangle at $T=0$ is the
ground state pressure~\protect\cite{ceperley87}.
(Data from the chemical model do not have a region of
negative $dP/dT$, since $r_s=2.0$ lies entirely in the
supercritical region.
The $dP/dT<0$ behavior does appear at higher densities.)
}
\end{figure}

As the temperature increases, dissociation occurs, as
is evident from the correlation functions shown in
Fig.~\ref{fig:grpp}.
Dissociation also results from isothermal compression.
As expected, $dn/dT < 0$ along the phase boundary,
due to cooperative thermal and pressure effects on the electrons.
While a first-order dissociation transition would proceed
isothermally at constant pressure, it occurs in a temperature
interval at constant volume.
At $r_s=1.86$, dissociation occurs for $6000 < T < 8000\,{\rm K}$, 
compared with $13000 < T < 15000\,{\rm K}$ in the chemical model~\cite{chabrier94:private}.
For Coulomb systems, the pressure is
$P=(n/3)[E+ K]$,
where $E$ and $K$ are, respectively, the total and kinetic energy
per atom.
Except at $r_s=2.2$, the pressure {\em decreases\/} during isochoric dissociation,
as shown in Fig.~\ref{fig:pressure} for $r_s=2$.
In the corresponding isobaric system, this unusual behavior becomes a positive
density discontinuity,
consistent with the negatively sloping phase boundary.
This is a strong indication of the presence of a first order
phase transition which terminates in a critical point near
$\left (r_s\approx 2.2, T\approx 11000\,{\rm K}\right)$.

\begin{figure}
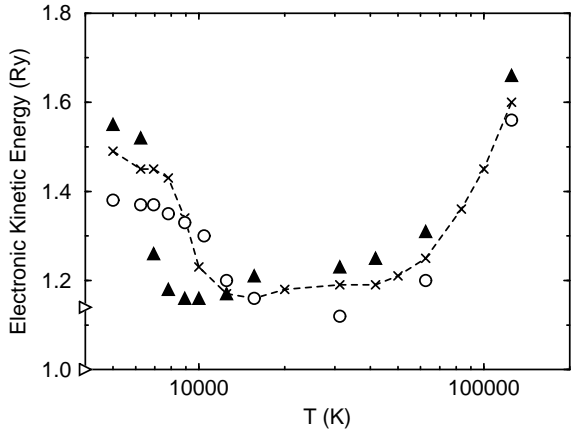

\epscenter{\figwidth}{fig3}
\caption
{
\label{fig:ke}
Electron kinetic energy per atom under isochoric cooling from
the plasma to molecular gas for three densities:
$r_s = 1.86$ (filled triangles), $r_s=2.0$ (crosses), and
$r_s = 2.2$ (open circles).
Statistical errors are smaller than the symbol size.
The kinetic energy increases as electrons pair to form bonds.
Different isochores cross, since compression suppresses molecular
formation.
The open triangles on the energy axis denote the kinetic energy
per atom for an isolated hydrogen atom and molecule, respectively.
}
\end{figure}

\begin{figure}
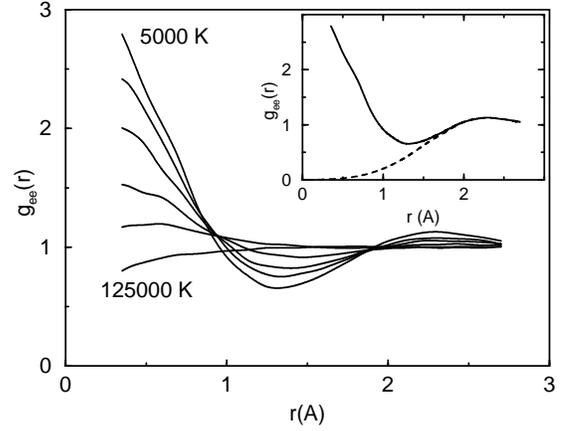

\epscenter{\figwidth}{fig4}
\caption
{
\label{fig:greeu}
Pair distribution functions, $g_{ee}(r)$, for spin-unlike electrons
at $r_s = 2.0$. 
Temperatures shown are 5000, 7813, 8927, 10000, 31250, and
125000 K.
Spin unlike electrons pair to form molecular ${\rm H}_2$ bonds.
Inset: spin-like (dashed) and spin-unlike (solid) distribution
functions at $r_s=2.0$, $T=5000\,{\rm K}$.
}
\end{figure}

A possible explanation for the existence of first order transition with
these behaviors lies in the increasing kinetic energy associated
with bond formation.
As shown in Fig.~\ref{fig:ke}, the electronic kinetic energy normally
decreases with decreasing temperature.
As molecules form, however, the kinetic energy increases.
The total kinetic energy per atom for an isolated atom and ${\rm H}_2$
molecule are also shown for comparison.
Since $g_{pe}(r)$ is nearly invariant during molecular formation,
the increase in kinetic energy results primarily from angular localization
as electrons leave spherical atomic-like states in favor 
of molecular bonding states.
Bond formation is clearly signaled by the pairing of spin-unlike
electrons, as shown in Fig.~\ref{fig:greeu}.
At low density, the increase in kinetic energy during bonding
is small relative to the total energy gain, so the pressure decreases.
As the density increases, however, the effects of the inter-molecular
repulsion begin to dominate, the molecules contract, and the
additional confinement of the electrons leads to an increasingly high
kinetic cost of binding.
Eventually, $\Delta P\propto\left [\Delta E + \Delta K\right ]$
changes sign, and the critical behavior appears.

To conclusively demonstrate the presence of this phase transition, we
must show that the above results persist in the limits $\tau\rightarrow 0$
and $N\rightarrow\infty$.
At $\tau^{-1} = 10^6\,k_B{\rm K}$, three-body and higher order terms in
the density matrix are significant in dense hydrogen.
Our pair-product density matrix slightly overestimates the probability
for two electrons to simultaneously occupy the area between a pair of
protons.
This leads to a slightly shortened bond length (by $0.01$ \AA) in $\rm{H}_2$,
overestimated kinetic, and correspondingly underestimated potential
energy, due to the enhanced electron-proton correlation.
Pressures, also affected, may be overestimated by as much as 15 GPa.
Finite size effects, quite large in the 32 electron ground state, are
substantially suppressed at finite temperature, due to smearing of
the Fermi surface.
Calculations of the free Fermi gas at $T=5000\,{\rm K}$ and $T=50000\,{\rm K}$
indicate the finite size errors are smaller than the
time-step errors.

It is interesting to consider the nature of the fluid into which the
dissociating hydrogen transforms.
This phase is not yet a plasma, as it retains very strong proton-electron
correlations.
It is perhaps best called a partially ionized atomic fluid, because the electron
binds to the proton sufficiently long to prevent other
electrons from approaching.
This behavior, apparent from $r^2 \left[g_{pe}(r)-1\right]$,
gradually disappears
as the temperature or pressure is raised and the fluid becomes fully ionized.
A more precise characterization of these electronic bound states and their eventual
disappearance can be made with the aid of
natural orbitals~\cite{girardeau90}, but this
requires a separate calculation of off-diagonal elements of $\rho(\beta)$.
We hope to perform this analysis in the future.

In conclusion, we have simulated dense hydrogen fluids by assembling a
fully interacting collection of protons and electrons.
The dissociation transformation, relevant to interior models of the
giant planets, occurs at somewhat lower temperatures than previous estimates.
We have identified in this system behaviors characteristic of
a first order phase transition.
The phase boundary has negative slope, due to the cooperative effects
of temperature and degeneracy in dissociation.
Molecular hydrogen dissociates not directly into a plasma, but first into a
partially ionized atomic fluid.

The spontaneous formation of molecules in this work,
in which the only inputs are the Hamiltonian 
and the nodal surface of a free fermion gas,
is an important and encouraging success for the restricted path-integral
Monte Carlo method for many-body Fermi systems.
All aspects of the method, including changing the trial nodes, can be
improved, so more accurate calculations are underway.
Although new calculations may alter these results, we
expect the basic findings to remain true.

A detailed tabulation of the hydrogen equation of state at many
experimentally inaccessible conditions, needed by planetary modelers,
will be straightforward to obtain.
Other possible applications include alkali metals, the
electron-hole liquid, and helium-hydrogen mixtures.
For the molecular fluid and solid at lower temperatures, the
errors due to free-particle nodes and time step error may
become substantial, so
more sophisticated nodes and high-temperature density
matrices will be needed before these cases can be studied.

This work was supported by the Office of Naval Research Grant No.
ONR N00014-92-J-1320, by the CNRS-NSF international travel
Grant No. NSF-INT-91-15682, and by the National Science Foundation
Grant No. NSF-ASC-94-04888.
The calculations were performed at the National Center for Supercomputing
Applications at the University of Illinois at Urbana-Champaign,
and the Cornell Theory Center at Cornell University.

%\bibliographystyle{prsty}
%\bibliography{../bibs/2dbose,%
%	../bibs/qmc,%
%	../bibs/hthy,%
%	../bibs/hexp,%
%	../bibs/pimc,%
%	../bibs/ppt,%
%	../bibs/magro,%
%	../bibs/books}

\end{document}